\documentclass[aps,prl,twocolumn,superscriptaddress,showpacs]{revtex4}

\usepackage{graphicx, latexsym, verbatim}
\usepackage{amssymb, amsmath}
\usepackage[ansinew]{inputenc}
\usepackage{overpic}
\usepackage{color}

\newcommand{\bEqa}{\begin{eqnarray}}
\newcommand{\eEqa}{\end{eqnarray}}

\begin{document}


\title{Scaling theory put into practice: first-principles modeling of transport
in doped silicon nanowires}



\author{Troels Markussen}
\affiliation{MIC - Department of Micro- and Nanotechnology, NanoDTU,
Technical University of Denmark, DK-2800 Kgs. Lyngby}

\author{Riccardo Rurali}
\affiliation{Departament d'Enginyeria Electrònica,
Universitat Autònoma de Barcelona 08193     , Spain}

\author{Antti-Pekka Jauho}
\affiliation{MIC - Department of Micro- and Nanotechnology, NanoDTU,
Technical University of Denmark, DK-2800 Kgs. Lyngby}

\author{Mads Brandbyge}
\affiliation{MIC - Department of Micro- and Nanotechnology, NanoDTU,
Technical University of Denmark, DK-2800 Kgs. Lyngby}


\date{\today}

\begin{abstract}
	We combine the ideas of scaling theory and universal conductance
	fluctuations with density-functional theory to analyze the
	conductance properties of doped silicon nanowires.
	Specifically, we study the cross-over from ballistic to diffusive
	transport in boron (B) or phosphorus (P) doped Si-nanowires by computing the mean free path, sample
	averaged conductance $\langle G\rangle$, and sample-to-sample
	variations std$(G)$ as a function of energy, doping density, wire
	length, and the radial dopant profile. Our main findings are: (i)
	the main trends can be predicted quantitatively based on the scattering properties
	of single dopants;
	(ii) the sample-to-sample fluctuations depend on energy
	but not on doping density, thereby displaying a degree of
	universality, and (iii) in the diffusive regime the analytical
	predictions of the DMPK theory are in good agreement with our ab
	initio calculations.
\end{abstract}

%


\maketitle Silicon nanowires (SiNWs) are strong candidates for
future nanoelectronic and sensor applications
\cite{LieberScience2001, CuiLieberNanoLett2003, LieberMatToday2005}. In
most of the demonstrated devices
the SiNWs are either
$p-$ or $n-$doped during the fabrication process. Very thin SiNWs
with diameters below 5 nm have been synthesized by several
groups \cite{DDDMa,HolmesScience2000,WuLieberNanoLett2004}, and due
to the small cross section, scattering by dopants is likely to be
very important. Moreover, due to reduced acoustic phonon scattering
in quasi one-dimensional systems, long coherence lengths might be
possible even at room temperature
\cite{LuLieberPNAS2005}, thus emphasizing the importance of defect
scattering. At the same time sample-to-sample
variations become a crucial issue: when the device length and the mean free
path are comparable, and shorter than
the coherence length, variations of the positions of the individual
dopant atoms can affect the conductance of the wire significantly.

The mathematical theory of the conductance of disordered quasi
one-dimensional systems has reached a high level of understanding in
the diffusive as well as the localization regime \cite{LeeRamakrishnan, Beenakker1997}.
A standard way to model disorder, analytically as well as
numerically, is to introduce random noise (Anderson disorder), which
in a tight-binding description is included through randomly varying
on-site energies. It does not seem obvious that a random disorder is
an adequate description of real physical disorder, such as dopants
or vacancies,  nor is there any obvious connection between a
physical defect density and the amplitude of the random disorder. In
particular, if only a few dopants are present in a wire, the system
is in the crossover region from ballistic to diffusive transport and
the discrete and local nature of the impurities must be modeled
adequately.

Several recent theoretical works considered dopants in SiNWs using density-functional
theory (DFT)
\cite{BlasePRL, PeelaersNanoLett2006, SinghNanoLett2006}, mainly
focusing on the structural and energetic properties of different
radial dopant positions. As an important first step towards the
modeling of physical  SiNWs, Fernandez-Serra et al.
\cite{BlaseNanoLett2006} considered  the scattering properties of
{\it single} P dopants in thin nanowires.

In this Letter we complete the analysis by calculating the conductance of {\it long} nanowires with
a {\it random distribution} of dopants (either P or B) along the
wire. We calculate the sample averaged conductance $\langle
G\rangle$, elastic mean free path (MFP) $l_e$, localization length $\xi$, and the sample-to-sample fluctuations characterized by
the sample standard deviation std($G$). We show that all these quantities can be understood and
accurately estimated from the scattering properties of the single dopants, implying that relatively
simple calculations are sufficient in practical device modeling.  The sample-to-sample fluctuations
at a given energy and dopant type vary with $L/l_e$ in a universal way independently on the dopant concentration,
and in the diffusive regime for wire length $l_e<L<\xi$, we observe good agreement with analytical
predictions of the Dorokhov-Mello-Pereyra-Kumar (DMPK) theory \cite{DMPK}.


{\it Method:} The length and energy dependent conductance of each sample
is found using the Landauer formalism together with a standard
recursive Green's function (GF) approach where the full scattering
region containing the dopant atoms is constructed by repeatedly
adding small unit cells \cite{Markussen2006}.
The unit cells are constructed using first--principles local orbital
DFT calculations \cite{siesta-ref,DFTnote}.  We emphasize that our
combined DFT and GF approach to calculate the conductance is fully
{\it ab initio} within the low bias coherent transport regime.

\begin{figure}[htb!]
\begin{overpic}[width=.48\textwidth]{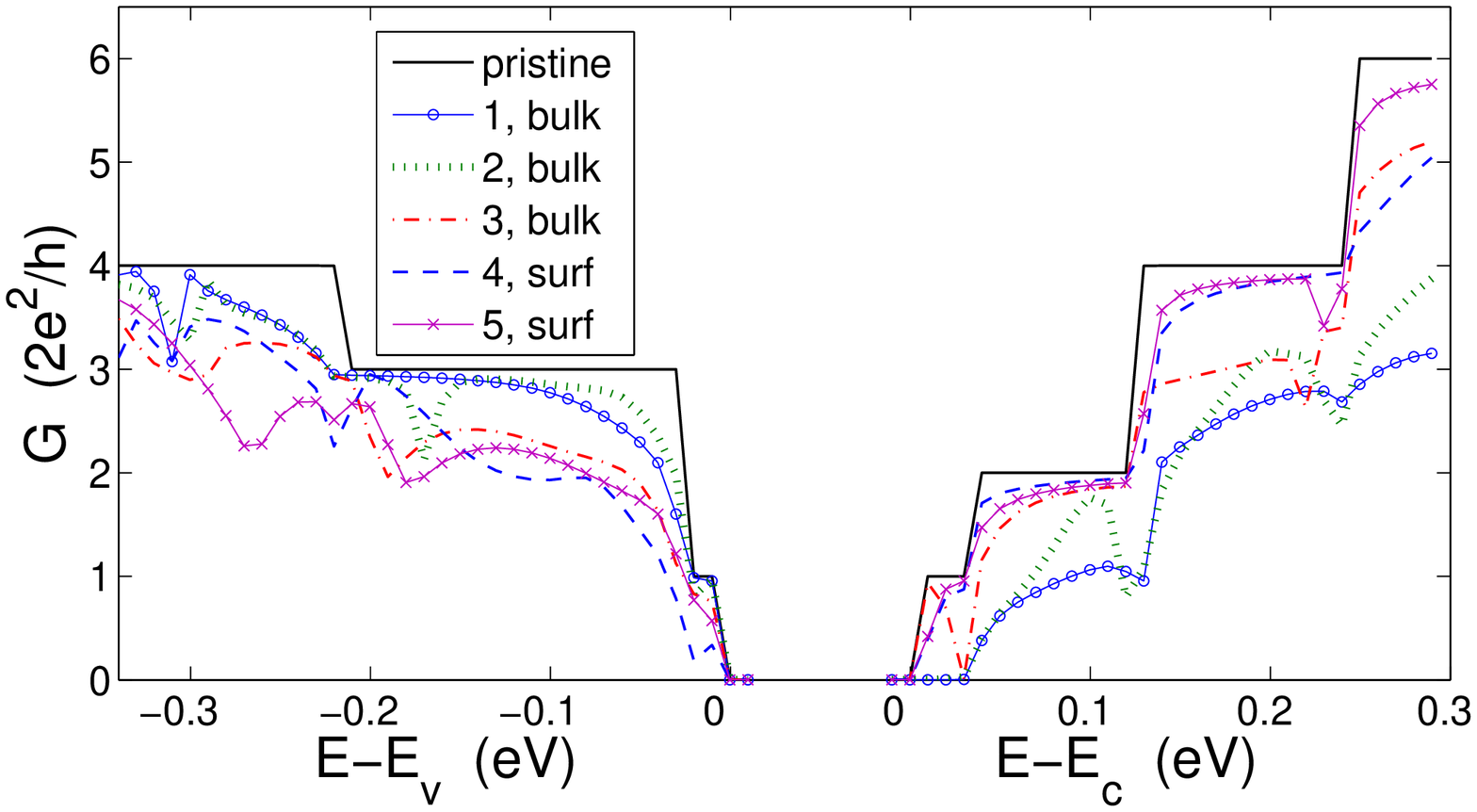}
    \put(47.,27){\includegraphics[width=.105\textwidth]{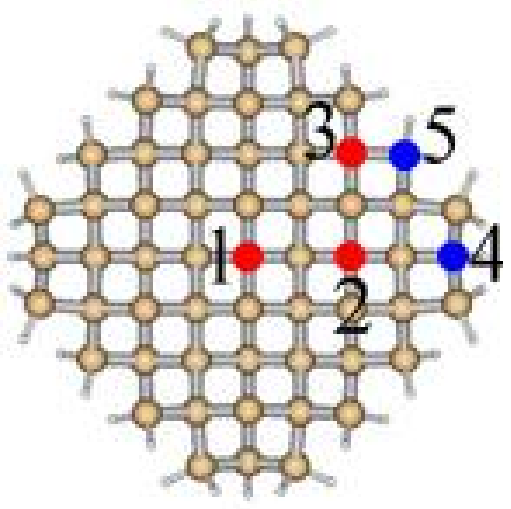}}
\end{overpic}
\caption{(color online). Transmissions through wires with a single B
(left) or P (right) dopant atom placed at the different radial
positions shown in the inset. Left: energies relative to the valence
band maximum, $E_v$; Right; energies relative to the conduction band
minimum, $E_c$. The band gap is $2.84\,$eV.}
 \label{ts-plots}
\end{figure}
{\it Single dopants:} Figure \ref{ts-plots} shows the transmission vs. energy through
infinite hydrogen passivated wires containing only a single dopant atom placed at five
different substitutional positions (1-5) indicated in the inset. The
left part shows the transmission for B dopants at energies in the
valence band while the right part shows results for P dopants at energies in the
conduction band. We notice that there are special resonant energies
where the conductance abruptly drops, similar to those observed in
Ref. \cite{BlaseNanoLett2006}, due to  enhanced local densities of
states at the dopant atoms associated with quasi-bound states. As also
pointed out in Ref. \cite{BlaseNanoLett2006}, there are significant
differences in the scattering properties between dopants located in
the bulk of the wire and those situated at the surface.
Interestingly, there are qualitative differences between P and B
dopants. While B in position 1 (middle of wire) is a weak scatterer
in the valence band, P at the same position is the strongest
scatterer in the conduction band.
For thicker wires the majority of the dopants are likely to be bulk-like and our
calculations thus predict P-doped wires to have smaller mobilities
than B-doped wires with the same dopant concentration, in agreement
with experimental results \cite{CuiLieberJPCB}.

{\it Long wires:} While the scattering properties of the single dopants are
interesting in their own right, real nanowires contain several
dopants with a certain distribution along the wire direction as well
as in the radial direction. Due to interference effects between
successive scattering events, it is not obvious if the single-dopant
results carry over to the long wire case. In the rest of this Letter
we investigate the conductance of long wires. In particular we
examine to what extent the long wire properties, such as mean
conductance $\langle G\rangle$ and variations std$(G)$, can be
determined from the scattering properties of the single dopants.

\begin{figure}[htb!]
\includegraphics[width=.48\textwidth]{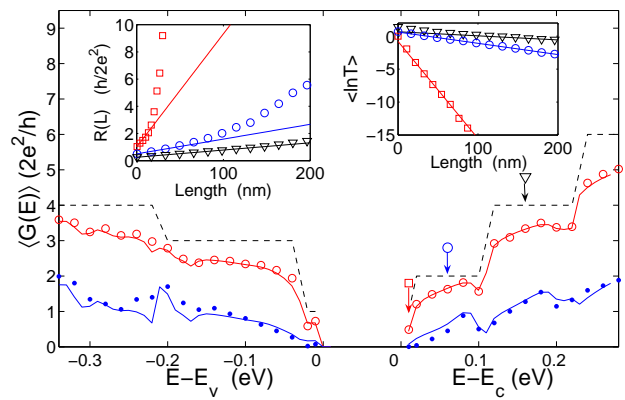}
    \caption{(color online). Energy dependent average conductance
    calculated at $L=10\,$nm (circles) and $L=100\,$nm (dots) with
    the GF method. Solid lines follow from Eq. \eqref{R_N}.
    Left inset: Resistance vs. wire length at energies $E-E_c=$ 0.01, 0.06 and
    0.16 eV
    (indicated with arrows in the main frame; same behavior is seen in the entire energy range).
    Solid lines are obtained using Eq. \eqref{R_N}. Right inset: $\langle \ln T\rangle$ vs.
    length at the same energies, where $T=G/(2e^2/h)$ is the transmission. Solid lines are
    linear fits in the interval 150 nm$\,<L<200\,$nm from which the localization length is determined.}
\label{G_vs_E_B_P}
\end{figure}

A single recursive GF calculation yields the conductance of a SiNW
with a given dopant distribution, length and energy. At each energy
we do this calculation for 300 different realizations of the dopant
positions, and we repeat these calculations for a range of wire
lengths, 10 nm $<L<$ 200 nm, all with the same dopant density. The
dopants are distributed randomly along the wire direction
\cite{RandomNote} as well as radially. The sample-averaged
resistance increases linearly for wire lengths shorter than the
localization length, as shown in the left inset in Fig.
\ref{G_vs_E_B_P}.
The initial linearly
increasing resistance defines the energy dependent MFP, $l_e(E)$,
through the relation $R(L,E)=R_c(E)+R_c(E)L/l_e(E)$ for wire lengths
$L<\xi$, where $R_c(E)=1/G_0(E)=h/(2e^2\,N)$ is the \textit{contact}
resistance of a pristine wire with $N$ conducting channels. It has
recently been shown that this definition of the MFP agrees with
values found using the Kubo formula and Fermi's golden rule
\cite{Avriller2006, Markussen2006}. In the linear resistance region,
we suggest that on average, the \textit{scattering} resistances from
the dopants add classically according to Ohm's law, i.e. the mean
resistance of a wire of length $L<\xi$ and average dopant-dopant
separation $d$ is given by
\begin{equation}
\langle R(L,E)\rangle=R_c(E) + \langle R_s(E)\rangle\,L/d. \label{R_N}
\end{equation}
$\langle R_s(E)\rangle$ is the average scattering resistance of the different dopant
positions which can be estimated from the single dopant
conductances, $G(E)$, in Fig. \ref{ts-plots} as $\langle
R_s(E)\rangle=1/\langle G(E)\rangle-1/G_0(E)$.  
From Eq. \eqref{R_N} we get an estimate of the MFP, $l_e'=R_c\,d/\langle R_s\rangle$.

Figure \ref{G_vs_E_B_P} shows the sample averaged energy dependent
conductance at wire lengths $L=10\,$nm (circles) and $L=100\,$nm
(dots) for B and P doped  wires
with an average dopant-dopant separation
$d=10\,$nm (corresponding to a bulk doping density of
$n\approx10^{19}\,$cm$^{-3}$) and uniform radial dopant
distribution. The dashed line shows the conductance of the pristine
wire, while the solid curves are estimated conductances obtained
from the single dopant transmissions shown in Fig. \ref{ts-plots}
using Eq. \eqref{R_N}. It is evident that the average conductances
are well reproduced by the single-dopant results.

We emphasize that Eq. \eqref{R_N} is only valid in the
quasi-ballistice ($L<l_e$) and diffusive ($l_e<L<\xi$) regimes. In
the localization regime ($L>\xi$), the resistance increases
exponentially, $R(L)\sim\exp(L/\xi)$. We have calculated the
localization length from the $\langle \ln T\rangle$ vs. L curves
shown in the right inset of Fig. \ref{G_vs_E_B_P} as $\xi=-1/{\rm
slope}$, and the slope is determined from a linear fit in the
interval $150\,$nm$<L<200\,$nm. In Table \ref{table1} we show the
resulting MFPs ($l_e$) and localization lengths ($\xi$) at four
different energies.
These are
compared with the estimates $l_e'=R_c\,d/\langle R_s\rangle$  based
on the single dopant transmissions, and $\xi'=l_e'\,(N+1)/2$ (this
relation follows from random matrix theory \cite{Beenakker1997}). It
is evident that both the MFP and the localization length can be
estimated fairly accurately from the single dopant transmissions
with a maximum error of $24\%$. We also note that the ratio
$\xi/l_e$ agrees with the prediction $(N+1)/2$ with a maximum
deviation of $30\%$.


\begin{table}[!htb]
\begin{center}
\begin{tabular}{ c|c| c| c| c| c}
  \hline
   \hline
  $E-E_c$ (eV) & $\;\;N\;\;$ & $\;\;l_e\;$ (nm) & $\;\;l_e'\;$ (nm) & $\;\;\xi\;$ (nm)& $\;\;\xi'\;$ (nm) \\[0.7mm]
  \hline
  0.01 & 1 & 10 & 8 &   7 &  8 \\[0.5mm]
  0.06 & 2 & 37 & 46 & 59 &  56 \\[0.5mm]
  0.16 & 4 & 49  & 47  & 133 &  123 \\[0.5mm]
  0.26 & 6 & 37  & 37  & 164  & 130  \\[0.5mm]
   \hline
   \hline
  \end{tabular}
  \caption{MFP and localization lengths
  obtained from sample averaging ($l_e$ and $\xi$),
  and estimated values, $l_e'$ and $\xi'$, obtained from the single dopant
  transmissions. $N$ is the number of conducting channels.}
  \label{table1}
  \end{center}
\end{table}

\begin{figure}[htb!]
\begin{center}
\includegraphics[width=.48\textwidth]{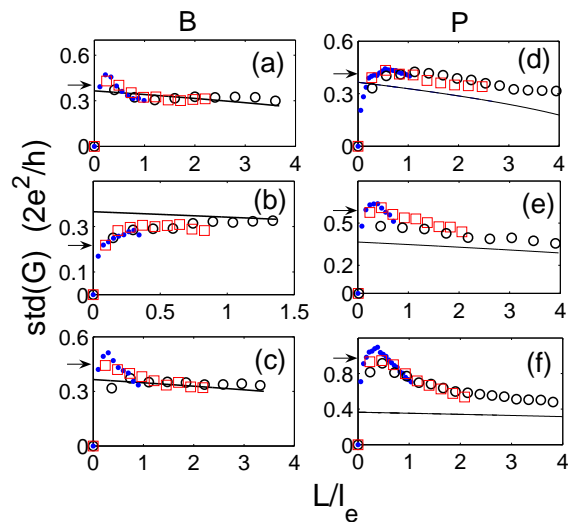}
\end{center}
\caption{(Color online) Sample standard deviation, std($G$) vs. normalized
length $L/l_e$ with average dopant-dopant separation $d=19\,$nm (dots),
$d=10\,$nm (squares) and $d=5\,$nm (circles).
(a)-(c) correspond to the energies $E-E_v=-0.12,\;-0.20,\;-0.30\,$eV in the
valence band for B doped wires, (d)-(f) correspond to the energies $E-E_c=0.06,\;0.16,\;0.26\,$eV
in the conduction band for P doped wires. The solid lines are analytical solutions to the
DMPK equation of Ref.\cite{Mirlin1994} which equals the UCF value $0.73\,e^2/h$ at $L=0$. The small arrows mark the maximum std($G$), $s$, estimated from the single-dopant transmissions.}
\label{stdG_vs_L_B_P}
\end{figure}

{\it Sample fluctuations:} An interesting question is whether or not the sample-to-sample variations can also be
understood in terms of the single-dopant transmissions. In the
diffusive regime we expect the variations to be close to the universal conductance fluctuation (UCF)
value, $0.73\,e^2/h$ for quasi-one dimensional
systems \cite{LeeStonePRL1985}.
Figure \ref{stdG_vs_L_B_P} shows the standard deviation, std($G$),
plotted against normalized length, $L/l_e$, for B (a)-(c) and P
(d)-(f) doped wires. 
The conductance fluctuations corresponding to different
concentrations lie close to each other, and we therefore conclude
that at a specific energy and type of dopant, the sample variations
are {\em independent} of dopant concentration but only depend on the
normalized length. This is in accordance with the theory of UCF
\cite{LeeStonePRL1985} and single parameter scaling theory
\cite{Abrahams1979} which predict that the sample fluctuations are
independent on the disorder strength. For modeling purposes this is
a very convenient result because one can limit the simulations to
only one dopant concentration.

In the diffusive limit, $L/l_e>1$, analytical results are known for
std$(G(L))$. For quasi one-dimensional systems with many conducting channels
and weak random disorder, the DMPK equation \cite{DMPK} predicts a weak length dependence,
std$(G(L))=\sqrt{8/15-64/315\cdot L/\xi}\,\cdot e^2/h$,
\cite{Mirlin1994}, shown in Fig.~\ref{stdG_vs_L_B_P} (solid lines).
As the length of the wire decreases, the sample fluctuations increase, and in most cases a
maximum in std($G$) is reached around $L/l_e=0.5$. In the limit $L\rightarrow 0$ there are no
dopants in the wire and the sample-to-sample variations vanish.

\begin{figure}[htb!]
\includegraphics[width=.48\textwidth]{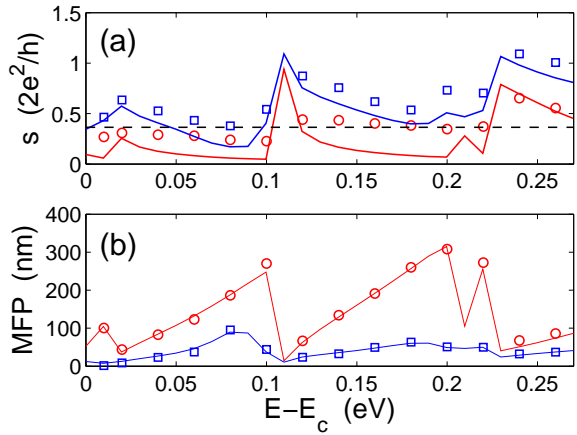}
    \caption{(color online). Maximal standard deviation (a)	 for for P doped wires ($d=10\,$nm,) with
    a homogeneous radial distribution (squares) and a pure surface doped wire (circles).
    The solid lines show the standard deviation, $s$, among the single dopant
    transmissions and the pristine wire, and the dashed line marks the UCF value $0.73\,e^2/h$.
    Panel (b) shows the MFP vs. energy for the two radial distributions
    (circles and squares). The lines represent values obtained from the single dopant transmissions. }
\label{MFP_vs_E_B_P}
\end{figure}

We next address this maximum value and the general behavior of the
sample-to-sample fluctuations vs. $L$ using the single-dopant
conductance results. Figure \ref{MFP_vs_E_B_P}(a) shows the maximum
values of std($G$) vs. energy for P-doped wires with a uniform
radial distribution (squares) and pure surface doping (circles). In
the former case all the dopant positions 1-5 are equally
probable\cite{radialNote} while in the latter the dopants can only
be in position 4 or 5, cf. inset in Fig. \ref{ts-plots}. The lower
solid line in Fig.\ref{MFP_vs_E_B_P}(a) shows the standard deviation
$s=$std($\{G_0,G_0,G_4,G_5\}$), where $G_0$ is the conductance of a
pristine wire while $G_4$ and $G_5$ are the single-dopant
conductances with the P dopant placed at position 4 and 5. This
sequence represents a situation where there is 50$\%$ chance for a
pristine wire, and 50$\%$ chance for a wire with a single dopant
either at position 4 or 5. The upper solid line in Fig.
\ref{MFP_vs_E_B_P}(a) shows similar values for the uniform dopant
distribution. When $s$ is larger than the UCF value (the horizontal
dashed line), the maximum std($G$) clearly follows the trends in
$s$. When $s$ is smaller than the UCF value, the maximum std($G$)
lies close to the UCF level.

We conclude that the shape of the std($G$) vs. $L/l_e$ curves in
Fig.~\ref{stdG_vs_L_B_P} can be qualitatively predicted from the
single-dopant transmissions: When $s$, indicated with small arrows
in Fig.~\ref{stdG_vs_L_B_P}, is smaller than the UCF value ($0.73
e^2/h$), we expect std($G$) to approach the analytical line from
below. The B doped wires in Fig. \ref{stdG_vs_L_B_P}(b) show such a
behaviour. Otherwise, if $s$ is larger than the UCF value, std($G$)
will reach a maximum value close to $s$ and approach the analytical
line from above.

{\it Mean free path:} Figure \ref{MFP_vs_E_B_P}(a) shows that the
sample fluctuations are significantly reduced in the surface doped
wires due to the weaker and more uniform scattering properties of
the surface positions. The MFP also depends significantly on the
radial distribution as seen in Fig. \ref{MFP_vs_E_B_P}(b), showing
the MFP for P doped wires with a uniform (circles) and pure surface
(squares) distribution of dopants. The solid lines are estimated
from the single-dopant transmissions using Eq. \eqref{R_N} once
again showing that the mean values of the conductance can be
accurately estimated from the single dopants. A very significant
increase in the MFP is observed for the surface doped wires as
compared to the uniform distribution. This might suggest that
increased device performance could be achieved if the P dopants are
located close to the surface, which indeed are the energetically
most favorable positions in the wires studied here and also found in
Ref. \cite{BlaseNanoLett2006, BlasePRL}. There are, however,
potential problems with dopants close to the surface, as they can be
passivated
by addition of an extra hydrogen atom \cite{BlaseNanoLett2006,SinghNanoLett2006}.

{\it In conclusion}, we have considered the conductance properties in B and P doped SiNWs.
We find that the sample averaged conductance and the sample-to-sample
fluctuations as well as the mean free path and localization length can be predicted quantitatively from the scattering properties of the single dopants.
Time consuming sample averaging can thus be avoided, which greatly
simplify modelling of the statistical conductance properties. These
findings may have a high impact on first-principles modelling of
electron transport in nanowires.


\begin{acknowledgements}
We acknowledge P. Marko\v{s} for useful comments. We thank the Danish Center for Scientific Computing (DCSC) for providing
computer resources. R.R. acknowledges financial support from Spain's Ministerio de Educaci\'{o}n y Ciencia Juan de la Cierva program and funding under Contract No. TEC2006-13731-C02-01.
\end{acknowledgements}


\begin{thebibliography}{24}
\bibitem[{Ran()}]{RandomNote}
\bibinfo{note}{Since the DFT calculations only contain single dopants the
  minimum dopant-dopant distance is 5 nm.}
\end{thebibliography}

\end{document}